\documentclass[12pt]{article}

\usepackage{graphicx}
\graphicspath{{./figures/}}
\usepackage{dcolumn}
\usepackage{bm}
\usepackage{amsfonts}
\usepackage{amsthm}
\usepackage{amsmath}
\usepackage{amssymb}
\usepackage{slashed}
\usepackage{color}
\usepackage{textcomp}
\usepackage{subfig}

\usepackage{hyperref}

\def\be{\begin{equation}}
\def\ee{\end{equation}}
\def\ba{\begin{eqnarray}}
\def\ea{\end{eqnarray}}

\def\bdm{\begin{displaymath}}
\def\edm{\end{displaymath}}
\def\la{~\mbox{\raisebox{-.6ex}{$\stackrel{<}{\sim}$}}~}
\def\ga{~\mbox{\raisebox{-.6ex}{$\stackrel{>}{\sim}$}}~}
\def\bq{\begin{quote}}
\def\eq{\end{quote}}

 at 10truept

\renewcommand{\[}{\left[}
\renewcommand{\]}{\right]}
\renewcommand{\(}{\left(}
\renewcommand{\)}{\right)}

\newcommand{\gam}{\gamma}

\renewcommand{\th}{\theta}

\newcommand{\bea}{\begin{eqnarray}}
\newcommand{\eea}{\end{eqnarray}}

\newcommand{\bi}{\begin{itemize}}
\newcommand{\ei}{\end{itemize}}

\newcommand{\beq}{\begin{equation}}
\newcommand{\eeq}{\end{equation}}
\newcommand{\beqa}{\begin{eqnarray}}
\newcommand{\eeqa}{\end{eqnarray}}

\def\la{~\mbox{\raisebox{-.6ex}{$\stackrel{<}{\sim}$}}~}
\def\ga{~\mbox{\raisebox{-.6ex}{$\stackrel{>}{\sim}$}}~}

\newcommand{\ov}{\overline}

\def\ltap{\ \raise.3ex\hbox{$<$\kern-.75em\lower1ex\hbox{$\sim$}}\ }
\def\gtap{\ \raise.3ex\hbox{$>$\kern-.75em\lower1ex\hbox{$\sim$}}\ }
\def\gl{\ \raise.5ex\hbox{$>$}\kern-.8em\lower.5ex\hbox{$<$}\ }
\def\roughly#1{\raise.3ex\hbox{$#1$\kern-.75em\lower1ex\hbox{$\sim$}}}

\begin{document}

\thispagestyle{empty}
\begin{flushright}
April 2018 \\
CERN-TH-2018-067
\end{flushright}
\vspace*{1.25cm}
\begin{center}
{\Large \bf Neutrino Masses from Outer Space}\\

\vspace*{1.25cm} {\large Guido D'Amico$^{a, }$\footnote{\tt
damico.guido@gmail.com}, Teresa Hamill$^{b, }$\footnote{\tt
teresahamill@gmail.com} and Nemanja Kaloper$^{b, }$\footnote{\tt
kaloper@physics.ucdavis.edu}}\\
\vspace{.3cm} {\em $^a$Theoretical Physics Department,
CERN, Geneva, Switzerland}\\
\vspace{.3cm}
{\em $^b$Department of Physics, University of
California, Davis, CA 95616, USA}\\

\vspace{1.5cm} ABSTRACT
\end{center}
Neutrinos can gain mass from coupling to an ultralight field in slow roll. When such a field is displaced from its minimum, its vev acts just like the Higgs vev in spontaneous symmetry breaking. Although these masses may eventually vanish, they do it over a very long time. The theory is technically natural, with the ultralight field-dependent part being the right-handed Majorana mass.
The mass variation induced by the field correlates with the cosmological evolution.
The change of the mass term changes the mixing matrix, and therefore
suppresses the fraction of sterile neutrinos at earlier times and increases it at later times.
Since the issue of quantum gravity corrections to field theories
with large field variations remains open, this framework may give an observational handle
on the Weak Gravity Conjecture.

\vfill \setcounter{page}{0} \setcounter{footnote}{0}
\newpage

Dark energy comprises over two thirds of the mass contents of the universe. It 
might be a cosmological constant. If so then it will affect the visible sector only by
gravitationally influencing the underlying geometry of the universe. If dark energy
is not constant, we can accommodate it in field theory as a quantum field with an
extremely flat potential, whose vacuum expectation value (vev) is displaced from the minimum. 
In the simplest realizations, the curvature of the potential, i.e. the mass of the field, needs to be smaller 
than the current Hubble parameter of the universe, $H_0 \sim 10^{-33} {\rm eV}$, in order 
for the field to remain suspended away from the minimum. In this regime 
the restoring force pulling the field to the minimum is overwhelmed by the
effective friction generated by the cosmic expansion. If this, so-called slow roll, regime can be
realized, the field will vary so slowly that it will imitate a nearly-constant dark energy. Commonly called 
{\it quintessence}, such ultralight fields by themselves 
do not provide a deep answer about the cosmological constant problem. In fact, if dark energy is 
really something like quintessence,  the problem would become even deeper, since in addition to
explaining the smallness of vacuum energy one would also need to explain the origin of the ultra-low scales
governing the quintessence sector. However, given our ignorance about the vacuum energy, some models
of quintessence \cite{fhsw,nomura,nilles,kalsor,kalsor2,trivedi} may at least serve the role of useful straw-men to guide observational searches. This is
all the more so since such fields might couple to other sectors non-gravitationally too. 

Generic quantum field theories that model such dynamics immediately run into problems.
A dark energy field needs to have a tiny shallow potential in order to sustain the present cosmic acceleration. So if it interacts directly with matter, it may induce a new long range force, competing with gravity. Also
the couplings would induce corrections to dark energy, which could disrupt slow roll. This is why we will focus here
only on a subset of quintessence models, which posses symmetries in the IR that can keep the corrections 
that could cause problems under control \cite{fhsw,nomura,nilles,kalsor,kalsor2,trivedi}. 

To evade these problems while keeping the couplings as large as can be, given by
Yukawa couplings\footnote{$\Gamma \propto 1$ or $\gamma_5$ for a scalar or pseudoscalar, respectively.}  of quintessence or its ultralight cousins to fermions, $\sim \frac{m_\psi}{M_{\rm Pl}} \phi \bar \psi \Gamma \psi$, requires very small fermion masses, much below the electroweak scale of the Standard Model (SM).
This only leaves neutrinos: they are massive but very light. Their loops would not destabilize ultralight bosons.
Therefore a quintessence field can directly couple to neutrinos and remain light enough to continue
behaving as dark energy even when quantum corrections are accounted for \cite{quintax}.  
Further, while coupled quintessence fields can mediate a long range force between neutrinos, the resulting bounds are feeble at present. Neutrinos are only a tiny fraction of the universe. In the early universe they would be much more significant. However if ultralight $\phi$ couple to $\nu$ only via Yukawas,
the extra force would be highly irrelevant,
being suppressed relative to gravity by the ratio $ m_\nu \bar \nu \Gamma \nu/T^\mu{}_\mu(\nu)$.

In turn, the direct Yukawa couplings of neutrinos to ultralight quintessence bosons lead to a novel source of neutrino mass terms. We will put these couplings in the
sterile neutrino Majorana mass, $m_S = g \phi$, $g \sim \frac{m_\nu}{M_{\rm Pl}}$, to preserve SM symmetries. When the ultralight field is displaced from the minimum of its potential and suspended in slow roll early on, the neutrino mass matrix will get contributions $\sim m_\nu \frac{\phi_0}{M_{\rm Pl}} \bar \nu \Gamma \nu \sim {\cal O}(1) m_\nu \bar \nu \Gamma \nu $.
The field vev $\phi_0$, and therefore also $m_S$, change slowly, with variation correlated with
cosmic evolution. The leading corrections to the boson potential are functions of $m_S$, the neutrino Dirac mass $m_D$, and the Higgs and $\phi$ Yukawas, and so the $\phi$ potential remains flat\footnote{Mechanisms for protecting quintessence from high energy physics, that the Higgs might be sensitive to, at higher loops are discussed in \cite{barbierihall,hill}.}. The variation of $m_S$ would change the mixing matrix, and adiabatically change the fraction of sterile neutrinos, suppressing it early on and increasing it later. This can alleviate
cosmological problems with having too many sterile neutrinos in earlier epochs. Moreover, in such a framework one can interpret the quintessence potential as arising solely form the sterile neutrino loops, after whatever
scale symmetry breaking, which gave neutrino the mass, occured \cite{barbierihall,hill}. This would also automatically explain why the quintessence field does not couple directly to any other SM fields, such as the Higgs: such couplings 
are prohibited by the low energy symmetries of the theory. 
In such models, $\phi$ was identified with the CP-violating phase of the $\nu$ mass matrix, and particle phenomenology and UV completions were discussed, however leaving out the variations of masses.
Here we will work with one active and one sterile neutrino, coupling the sterile to $\phi$ which
needn't be a phase of the $\nu$ mass matrix. Our analysis applies to quintessence, but also
to other ultralight $\phi$s, which may arise in ultralight dark sector monodromies\footnote{Interesting models involving couplings to bosons which, while light, are much heavier than those we consider, and can be fuzzy dark matter, were discussed, for example, in \cite{DM}.} \cite{quintax,jaeckel}. We stress that in this work we are not trying to explain the origin of the quintessence field and its ultra-low parameters. We are merely aiming to use the quintessence field
as a contributing sector to the neutrino mass, which would lead to interesting extremely slow, but observable, time variation of the neutrino masses and mixings. 

The required field displacement from the minimum, ${\cal O}(M_{\rm Pl})$, can be accomplished
in field theory limit by protecting the $\phi$ sector with symmetries, and there
is ample literature on the subject \cite{fhsw,nomura,nilles,kalsor,kalsor2,trivedi}. However UV completing such theories in quantum gravity is  less
clear. There are examples of difficulties with embeddings into string theory, where field ranges appear to be limited
to near-sub Planckian scales \cite{banksdine,witten}, and concerns that Weak Gravity Conjecture (WGC) \cite{nima} might provide bounds that
preclude long slow roll. On the other hand, the observations are pushing in the other direction: if dark energy is a field in slow roll, it must range by at least mildly super-Planckian scales to get
the equation of state sufficiently close to $-1$ \cite{qbounds}. This tension raises an interesting prospect of using neutrino cosmology to test quantum gravity. If neutrino masses, sterile neutrino fraction and dark energy vary in a correlated way, Planckian fields may be a cause. This could force rethinking aspects of WGC. While the opposite would have less impact, clearly the neutrino mass variation might yield a novel probe of the slow roll regime and the dark energy sector of the universe.

Neutrinos in the minimal SM are left-handed Weyl spinors $\nu_L$ (with  their CP partners $\nu^c_R = C \ov{\nu_L}^T$), partnered in an SU(2) doublet with a charged lepton.
Since their weak isospin and hypercharge are $1/2,-1$, respectively,
one cannot write relevant mass terms for them alone. To introduce the mass terms, therefore, one must add
right-handed singlet (or sterile) neutrinos $\nu_R$ (with $CP$ partner $\nu^c_L = C \bar{ \nu_R}{}^T$).
We can write two mass terms: a Dirac mass, mixing the left- and right-handed $\nu$s
and involving Yukawa coupled Higgs to preserve gauge symmetry, and Majorana mass for the
right-handed singlet. In unitary gauge, the mass terms are
\be
 y_{\nu} H \( \ov{\nu_L} \nu_R + \ov{\nu_R} \nu_L \) + \frac{m_S}{2} \( \ov{\nu^c_L} \nu_R + \ov{\nu_R} \nu^c_L \) =
 y_\nu (v + h) \ov{\nu_D} \nu_D + \frac{m_S}{2} \ov{\nu_M} \nu_M \, .
 \label{num}
\ee
We use the Dirac field $\nu_D = \nu_L + \nu_R$ and the Majorana 4-spinor $\nu_M = \nu^c_L + \nu_R$. Here $v$ is the Higgs vev, and $h$ the physical scalar and the Yukawa coupling is $y_{\nu} \simeq 10^{-12}$, taking into account that $v = {\cal O}(100) \, {\rm GeV}$. The Dirac $\nu$ mass is $m_D = y_\nu v$, while Majorana mass $m_S$ is a free parameter. Theories of neutrino masses exploit this freedom to explain the origin of the mass scales, usually resorting to UV physics to derive $m_S$. However, since $\nu_R$ is a fermion singlet, a small $m_S$ is technically natural in the sense that $m_S \rightarrow 0$ is protected by chiral symmetry of the theory. For this reason, we can imagine that $m_S$ originates from IR physics, and try to generate it by couplings to an ultralight boson.

If $\phi$ is either an SM singlet scalar $\phi_S$ or a pseudoscalar $\phi_A$, its couplings to $\nu$s are
\be
\frac{g_{\rm s}}{2} \phi_S \ov{\nu_M} \nu_M + i \frac{g_{\rm a}}{2} \phi_A \, \ov{\nu_M} \gam_5 \nu_M
= \frac{g_{\rm s}}{2} \phi_S \[ \ov{\nu^c_L} \nu_R + \ov{\nu_R} \nu^c_L \] +  i \frac{g_{\rm a}}{2} \phi_A \[ \ov{\nu^c_L} \nu_R - \ov{\nu_R} \nu^c_L \] \, .
\label{qnu}
\ee
Since $\phi$ is a SM singlet it can couple to the Higgs (squared) field via renormalizable couplings. Generic couplings would yield a large mass for $\phi$. These must be absent at the tree level. The theory must start with only sterile
$\nu$ directly coupled to $\phi$, which we impose directly\footnote{This seems to us to be less implausible now, at the dawn of the post-naturalness era \cite{judge}.}. The Higgs-$\phi$ couplings are induced by loop corrections, but they will be weak for small $m_\nu$.

When $\phi$s are slowly evolving\footnote{The environmental population of sterile neutrinos can induce a $\phi$ drift during radiation epoch when $\rho_\nu$ is larger than $V(\phi)$. The maximal variation due to this can be shown to be $\Delta \phi < \frac{g M_{\rm Pl}}{m_\nu} \frac{m_\nu}{T_{eq}} M_{\rm Pl} \ll M_{\rm Pl}$, where $T_{eq}$ is the temperature at radiation-matter equality, and so we ignore it in what follows.}, with a large initial vev $\phi_0$, possibly $\sim M_{\rm Pl}$, they induce effective Majorana masses. Indeed, the scalar term in (\ref{qnu}) reduces to $\frac{g_{\rm s} \phi_{S0}}{2}  \[ \ov{\nu^c_L} \nu_R + \ov{\nu_R} \nu^c_L \]$, precisely the Majorana mass in (\ref{num}). The pseudoscalar mass  $i\frac{g_{\rm a} \phi_{A0}}{2}  \[ \ov{\nu^c_L} \nu_R - \ov{\nu_R} \nu^c_L \]$ can be put in this form by a chiral rotation; when both terms in (\ref{qnu}) are present, a more general chiral rotation will bring the mass term to $\frac{1}{2} (g_{\rm s}^2 \phi_{S0}^2+ g_{\rm a}^2\phi_{A0}^2)^{1/2}$. In what follows we will ignore the difference\footnote{Strictly speaking, the difference between $\phi_S$ and $\phi_A$ can only arise when the fermion is massive, and chiral symmetry is broken.} between $\phi_S$ and
$\phi_A$, and use $g$ for either $g_{\rm s}, g_{\rm a}$. If both scalars are present with comparable couplings, the leading results may only differ by ${\cal O}(1)$ terms. We will focus on $m_S > m_D$, where the $\phi$ effects are maximized.

Now we can consider loop corrections to the theory. It is useful to transition to propagation eigenstates
of the neutrino mass matrix (\ref{num}).
Since the left-handed Majorana mass vanishes, we can make $m_D, m_S$ real. The mass eigenstates
are $n_\pm \equiv n_{\pm L} + n^c_{\pm R}$, where $n_{\pm k}$ are obtained from the interaction eigenstates
via a unitary rotation, with the mixing angle defined by $\tan (2\theta) = 2 m_D/ m_S$, and the eigenmasses $m_\pm = \frac{1}{2} ( \sqrt{m_S^2 + 4 m_D^2} \pm m_S )$:
\be
\begin{pmatrix}
    \nu_L \\ \nu^c_L
\end{pmatrix}
=
U
\begin{pmatrix}
    n_{-L} \\ n_{+L}
\end{pmatrix} \, ,
\qquad
\begin{pmatrix}
    \nu^c_R \\ \nu_R
\end{pmatrix}
=
U^\text{*}
\begin{pmatrix}
    n^c_{-R} \\ n^c_{+R}
\end{pmatrix} \, , \qquad {\rm where} ~~
U =
\begin{pmatrix}
    i \cos \th & \sin \th \\
    -i \sin \th & \cos \th
\end{pmatrix} \, .
\label{matrices}
\ee
We stress that due to the simplicity of our setup,
$\theta$ and $m_D/m_S$ are {\it not} independent, but correlated. In more general frameworks they would be independent variables.

Rewriting the Lagrangian involving (\ref{qnu}) in terms of $n_\pm$, we use the background field method to compute the one-loop effective action for $\phi$ by splitting it and the Higgs into $\phi_0 + \varphi$, $H = v + h$, absorbing $\phi_0$ into $m_S = g \phi_0$ (valid over times
shorter than the Hubble time ${\cal H}^{-1}$) and then integrating out $\varphi$, $\nu$ and $h$ in
\be
Z = e^{iS_0(\phi_0) + i\int d^4x \phi_0 J} \int \bigl[{\cal D}\varphi {\cal D} \nu {\cal D} h \bigr] \exp\Bigl\{i \Bigl( \frac{1}{2} \int d^4x d^4y \sum_{n,l} \frac{\delta^2 {\cal L}(\phi_0) }{\delta \varphi_n(x) \delta \varphi_l(y)}\varphi_n(x) \varphi_l(y) + \ldots \Bigr)\Bigr\}  \, .
\ee
The sum is over $\varphi_l = (\varphi,h)$. This functional determinant can be evaluated in perturbation theory by summing the leading
renormalized one-loop diagrams, such as those given in Fig. 1, as dictated by the SM gauge symmetry and lepton couplings.

\begin{figure*}[thb]
\centering
\includegraphics[scale=0.36]{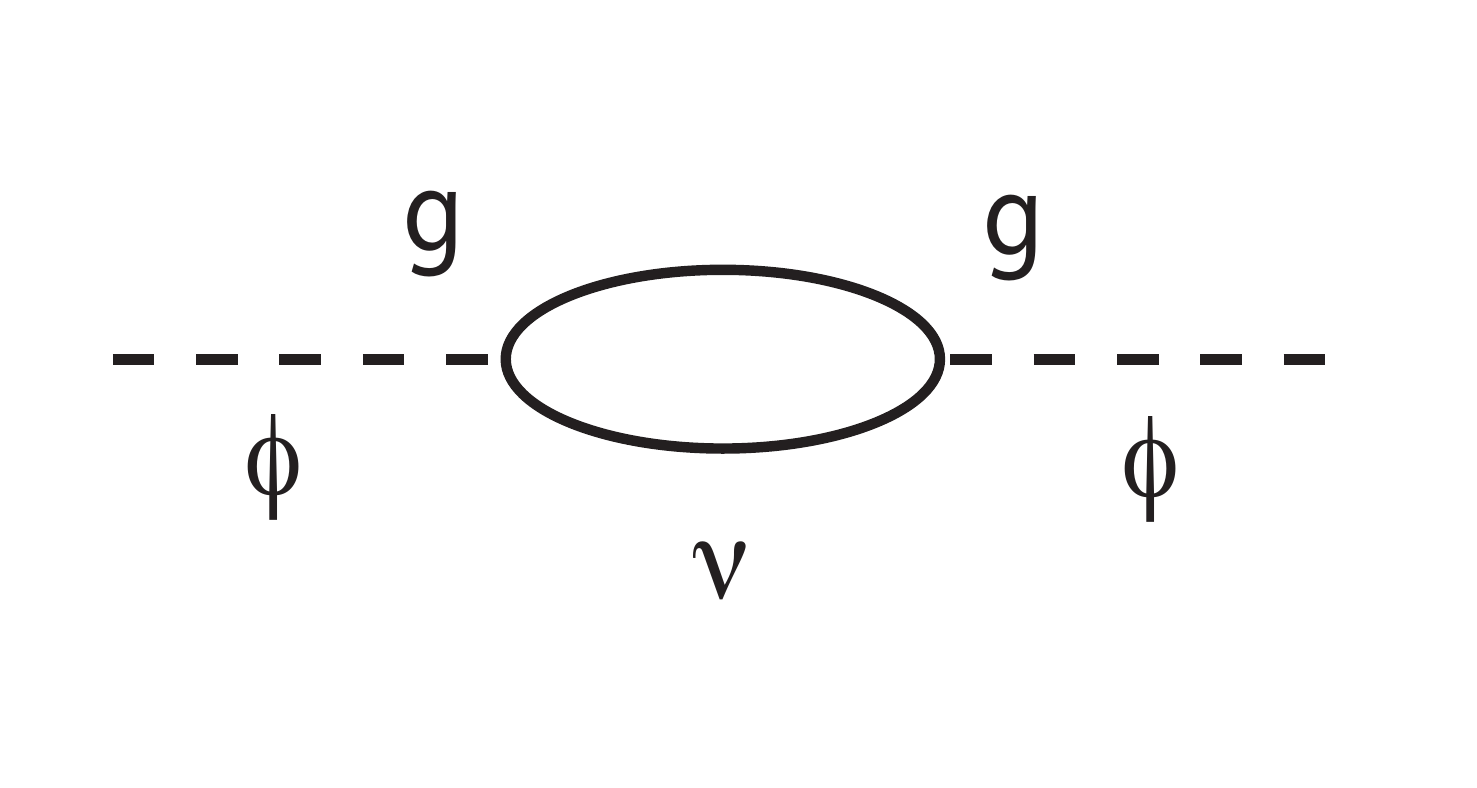}
\includegraphics[scale=0.36]{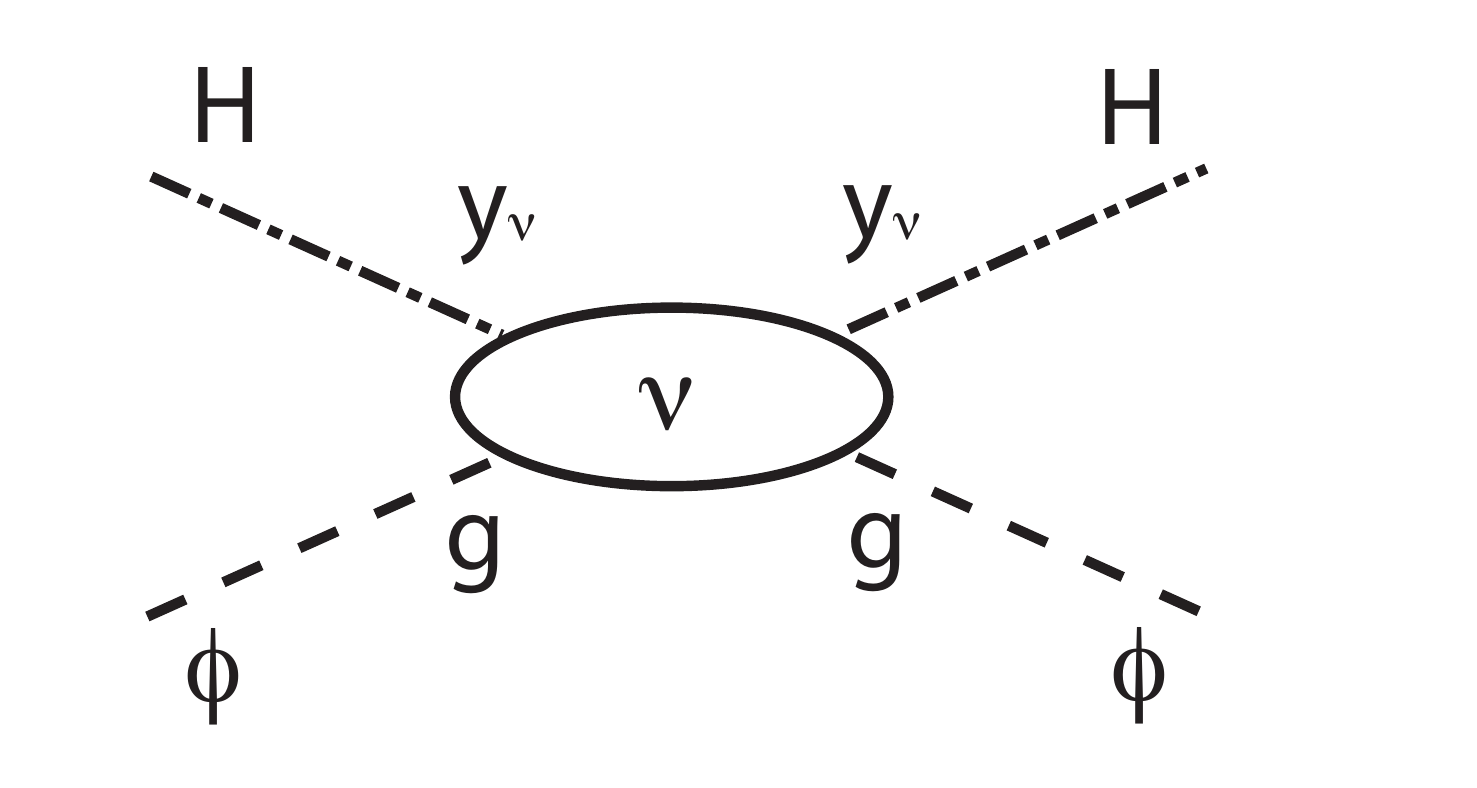}
\includegraphics[scale=0.35]{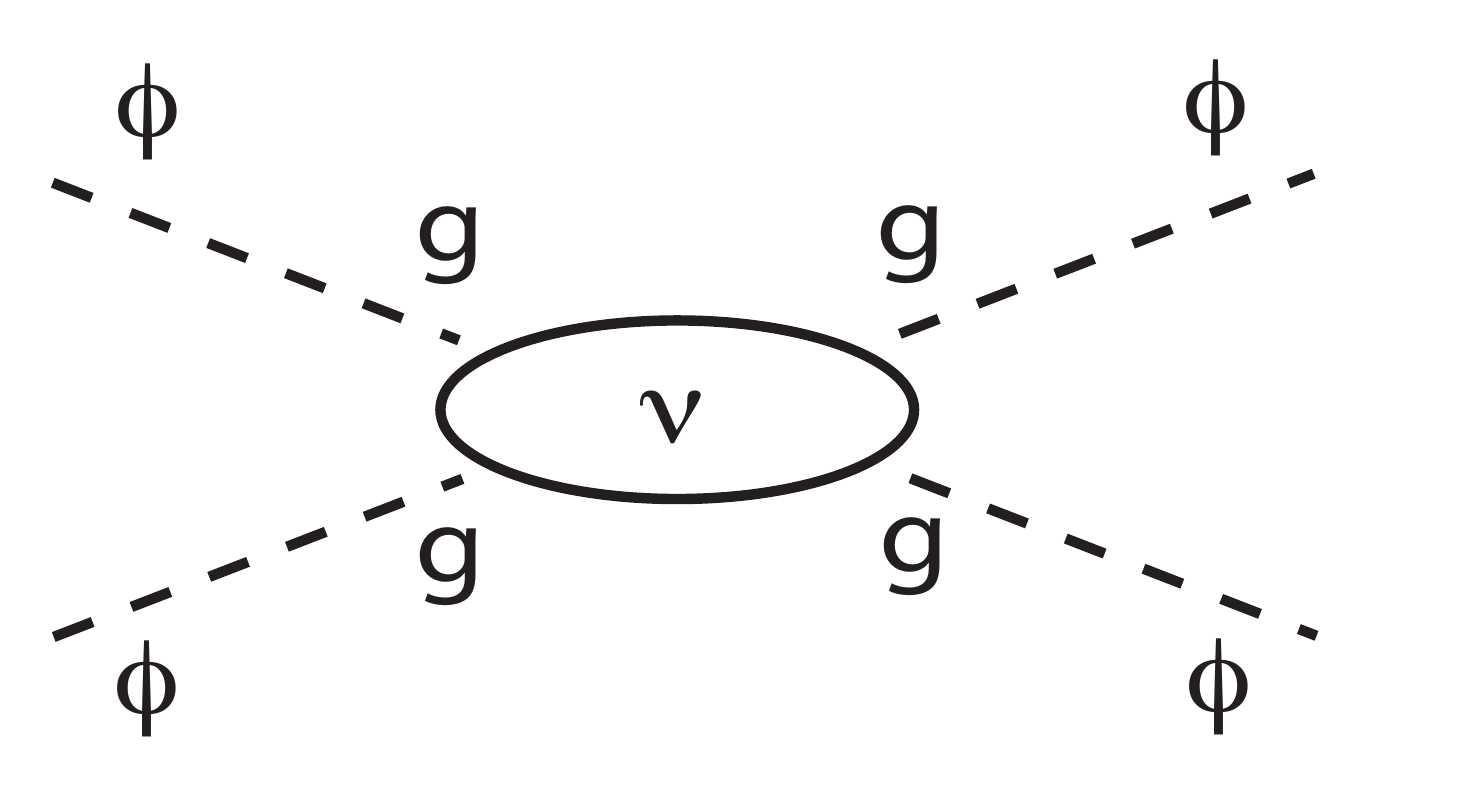}
\caption{Examples of neutrino loop contributions to the $\phi$ potential}
\label{fig1}
\end{figure*}
The leading corrections to $\phi$ potential are
\be
\delta V = \frac12 \delta m^2 \phi^2
+ \frac{\lambda}{4} |H|^2 \phi^2 + \frac{\lambda'}{4!} \phi^4 \, .
\label{potcorr}
\ee
The full one loop potential can be resummed in the Coleman-Weinberg form, but for our purposes the leading order
terms  (\ref{potcorr}) are sufficient to show that the coupling does not spoil the flatness of the quintessence potential. 
The parameters are given by linear combinations of the powers of $\nu$ masses, the mixing angle terms and couplings as per the diagrams in Figure 1. E.g., $\delta m^2$
up to terms which are independent of the logs of the renormalization point (that can be renormalized away) are
\be
\delta m^2 =  \frac{ g^2}{16\pi^2} \Bigl(\sin^4 \theta \, m_-^2 \ln(\frac{\mu}{m_-}) + \cos^4 \theta\, m_+^2\ln(\frac{\mu}{m_+}) + \sin^2 \theta \cos^2 \theta \, \frac{(m_-^3 \ln(\frac{\mu}{m_-}) + m_+^3 \ln(\frac{\mu}{m_+}) ) }{3(m_-+m_+)} \Bigr) \, .
\ee
The formulae for $\lambda$ and $\lambda'$ involve similar combinations of neutrino masses, couplings and mixing angles,
dictated by dimensionality. Again up to log-independent terms,
\be
\lambda =  \frac{g^2 y_\nu^2}{16 \pi^2} \Bigl({\cal O}(1) \ln(\frac{\mu}{m_-}) + {\cal O}(1) \ln(\frac{\mu}{m_+}) \Bigr) \, ,
~~~~~~~~~~  \lambda' \sim \frac{g^4}{16\pi^2} {\cal O}(1) \ln(\mu/m_\pm) \, .
\ee

If $\phi$ is to be quintessence, its mass must be $m\la {\cal H}_0 \sim 10^{-33} {\rm eV}$. For $V(\phi)$ to remain
that flat, $\delta m^2 < m^2 \la (10^{-33} {\rm eV})^2$, and so
$g m_\pm \la 4\pi  \times 10^{-33} {\rm eV}$. To ensure that the quadratic is small enough, $\lambda' \la 10^{-121}$ since $\phi \ga M_{\rm Pl}$, we need $g\la 10^{-30}$. Hence $m_\pm \la 4\pi \times 10^{-3} {\rm eV}$.
Since the gravitational potential $\propto m_\pm/M_{\rm Pl}$, the long range force between neutrinos mediated by quintessence is at most comparable to gravity.
All this shows that $\lambda$ is very small too, $\lambda \la 10^{-86}$: the Higgs mass shift due to the biquadratic  is tiny when $\phi \ga M_{\rm Pl}$, and the Higgs vev shifts the $\phi$ mass by $\la {\cal H}_0$.

If $\phi$ is heavier, the numbers scale up accordingly. If $\phi$ is an ultralight field which arises in monodromy constructions of quintessence \cite{quintax}, its mass could easily be several orders of magnitude larger than
$10^{-33} {\rm eV}$. Such a component of the universe could be as much as few percent of the total critical
energy density \cite{amendolabarbieri}. Such fields typically start with large initial vevs $\sim f$, where $f \la M_{\rm Pl}$ is their decay constant, and stay in slow roll for a long time. Thus $g$ could be larger than $10^{-30}$, but not too large or it would overclose the universe. As a result,
$m_\pm \la m/g$ could be greater than milli-eV, but not arbitrarily large. So an ultralight boson heavier than $10^{-33} \,{\rm eV}$ could consistently couple to neutrinos with masses $\sim {\rm eV}$.

This analysis shows that the interesting regime of parameters is $m_D < m_S \la m/g$, guaranteeing stability
of the ultralight boson sector, and making the mechanism phenomenologically interesting. In principle
$m_S < m_D \la m/g$ is consistent with radiative stability, but this, the pseudo-Dirac, regime (see \cite{langacker} for terminology) is both less interesting phenomenologically and more constrained given the scales of our cosmological mass terms.
With $m_D < m_S$ we can imagine two regimes, $m_D \la m_S$ and $m_D \ll m_S$. The mixing regime $m_D \la m_S$ is particularly interesting since Dirac and Majorana mass remain comparable.  The problem with this regime is that the propagation eigenstates involve significant contributions from both active and sterile neutrinos and may lead to sterile neutrinos equilibration. This could run afoul of the cosmological bounds which limit the effective number of neutrino species \cite{barbieridolgov,Langacker:1989sv,murayama,strumia,hannes}, which has prompted a lot of work on sterile neutrinos \cite{white}. In our case, since $m_S$ changes as $\phi$ rolls, the sterile neutrino -- being heavier early on, with larger $m_S$ -- may be sufficiently decoupled originally. It only starts to couple more significantly as $m_S$ decreases, and the mixing angle
$\theta = \frac12 \tan^{-1}(2m_D/m_S)$ grows.

This trend can be further enhanced if $\nu$ couples to $\phi$ which is heavier than quintessence.
As we noted, such fields might comprise as much as few percent of current critical energy density of the universe \cite{amendolabarbieri,quintax}. Initially we can have $m_D \ll m_S$, so that the see-saw regime completely decouples the sterile neutrino: the (mostly active) neutrino mass is $m_- \simeq m_D^2/m_S$ and it is much smaller than the (mostly sterile) neutrino mass $m_+ \sim m_S$. Since initially $\phi$ is in slow roll this would govern the neutrino sector until a very late time when ${\cal H}$ drops below the $\phi$ mass. At this point $\phi$ falls out of slow roll and starts to oscillate (slowly, with a period $\sim m$, corresponding to tens of millions of years or more!) around the minimum. The amplitude of $\phi$ would dilute
by the expansion of the universe as $\sim 1/a^{3/2}$. This can reduce $\phi$, and so also $m_S$, significantly, by as much as $\sim 10^5$, changing the neutrino
mixing matrix from the see-saw regime to the mixed regime, where the masses of the mostly active and mostly sterile neutrino are comparable, $m_- \sim m_+$ (see (\ref{matrices})).

As a consequence, the mass evolution from heavy to light will suppress the sterile $\nu$ population in the early universe, satisfying the cosmological bounds \cite{barbieridolgov,Langacker:1989sv,murayama,strumia}, while enhancing the active-sterile mixings at present. Indeed, let us compare the sterile $\nu$ abundances in the early universe for two different values of $m_S$ and $\theta$, using (\ref{matrices}). Since the sterile $\nu$ is a singlet, it
will be generated only by the flavor oscillations with the active $\nu$.  We take the transformations (\ref{matrices}) and
impose the initial condition $\nu_L \sim \nu^c_R \ne 0$, by equilibration with the other SM degrees of freedom, and
$\nu_R = \nu_L^c=0$. The average population of sterile neutrinos after a few ``oscillation times", which are much shorter than ${\cal H}^{-1}$, will be $\sim N_+ \sim \sin^2(\theta) N_{initial}$. Here
$N_+ \sim |n_+|^2$ is the population of heavier, mostly sterile $\nu$s. Thus the populations with different $\theta$ are related by
\be
\frac{N_+^{(1)}}{\sin^2(\theta_1)} = \frac{N_+^{(2)}}{\sin^2(\theta_2)} \, .
\label{nupop}
\ee
We can use (\ref{nupop}) to estimate the $\nu$ fractions in the same theory where $m_S = g \phi_0(t)$ adiabatically changes in time, over time scales ${\cal H}^{-1} \gg 1/m_\pm$. If $m_S \gg m_D$ early on,
after equilibration the neutrinos are mostly active, with number density $N^{(1)}_- \sim |n_-|^2 \propto N_{initial}$,
and
\be
N_S^{(1)} \sim N^{(1)}_+ \propto \sin^2(\theta_1) N_{initial} \sim \frac{m_D^2}{ m_S^2(t_1)}  \, N_{initial} \, .
\ee
Later on, as $m_S$ decreases toward $m_D$, the active neutrinos oscillate into sterile ones, and
\be
N^{(2)}_S \sim N_-^{(2)} \sim N_+^{(2)} \sim \sin^2( \theta_2) \, N_{initial}  \, \Bigl(\frac{a(t_1)}{a(t_2)}\Bigr)^3  \sim \frac{m_D^2}{ m_S^2(t_2)}  \, N_{initial} \,  \Bigl(\frac{a(t_1)}{a(t_2)}\Bigr)^3  \, ,
\label{numdens}
\ee
where $\Bigl(\frac{a(t_1)}{a(t_2)}\Bigr)^3$ accounts for the cosmological redshift factors.
Thus the initial fraction of sterile neutrinos is suppressed relative to the late one by
$(m_S(t_2)/m_S(t_1))^2$ -- which is just {\it decoupling}. Notice
that even if $m_S$ changes by only a factor of few initially, the fraction of mostly sterile neutrinos is suppressed at earlier times by at least
an order of magnitude. Since at late times the mostly sterile neutrinos couple more strongly, being lighter, they could be detectable through stronger mixings in lab experiments. The parameters of the model would be constrained by the experiments, but
in order to study their detailed effects on our proposal we'd have to generalize the model to the full three families.

Time variation of neutrino masses correlates with the cosmological evolution of $\phi$,
\be
\frac{\dot m_\pm}{m_\pm} = \pm \, \frac{m_+-m_-}{m_++m_-} \, \frac{\dot \phi}{\phi} \, .
\label{massvar}
\ee
For a field in slow roll, where $1+ w_{DE} \simeq \dot \phi^2/V_{DE}$, $\dot \phi/\phi \simeq - \sqrt{3 \Omega_{DE}(1+w_{DE})} M_{\rm Pl} {\cal H}_0/\phi$. On the other hand, if $\phi$ is oscillating around its minimum, then by virial theorem $\dot \phi/\phi \simeq - m$. Thus
\be
\frac{\dot m_\pm}{m_\pm} \simeq
\begin{cases}
\mp \, \frac{m_+-m_-}{m_++m_-} \, \sqrt{3 \Omega_{DE}(1+w_{DE})} \, \frac{ M_{\rm Pl}}{\phi} \, {\cal H}_0 \, , ~~~~~~ {\rm for} ~~ m < {\cal H}_0 \, ; \\
\mp \, \frac{m_+-m_-}{m_++m_-}  \, m  \, , ~~~~~~~~~~~~~~~~~~~~~~~~~~~~~~~~~~~~ \, {\rm for}~~  m > {\cal H}_0 \, .
\end{cases} \label{massamplitude}
\ee
These rates of change are far too small to be seen in lab experiments, but they may be detectable by cosmological observations. Further, in the latter case where $\nu$s couple to a $\phi$ heavier than quintessence, it is natural to have $\phi$ also couple to quintessence \cite{quintax}. Hence we could end up with a system of dark energy coupled to ultralight dark matter which in turns couples to neutrinos. This may have interesting cosmological implications and warrants further study, but is beyond the scope of the present work.

Because $\phi$ is so light, its exchange will mediate a new long range force between neutrinos, as given by the
Feynman diagram of Fig. 2. If $\phi$ is a scalar, the force could go as $\propto 1/r^2$. In contrast to \cite{mavans}, the faster clustering of $\nu$s driven by this force at scales
$\ell \la 1/m$ \cite{zaldarriaga} is not a problem since now the background neutrino density needn't be homogeneous to support cosmic acceleration. If $\phi$ is dark energy, it is in slow roll by itself. Moreover, since
$m$ is much smaller than in \cite{mavans} the clustering time scales are much longer, and might be as long as the age of the universe - but certainly not much shorter than millions of years. Such structures would be much slower to form and much less dense. For a pseudoscalar $\phi$ the extra forces would drop off faster at large distances \cite{moodywilczek}, and would not present cosmological problems. Either way the forces would not be much
stronger than gravity and they are not constrained significantly by laboratory neutrino experiments \cite{spin}. The smallness of $g$ also guarantees that the bounds from supernovae emissions and stellar cooling are satisfied \cite{stars}.
\begin{figure*}[thb]
\centering
\includegraphics[scale=0.35]{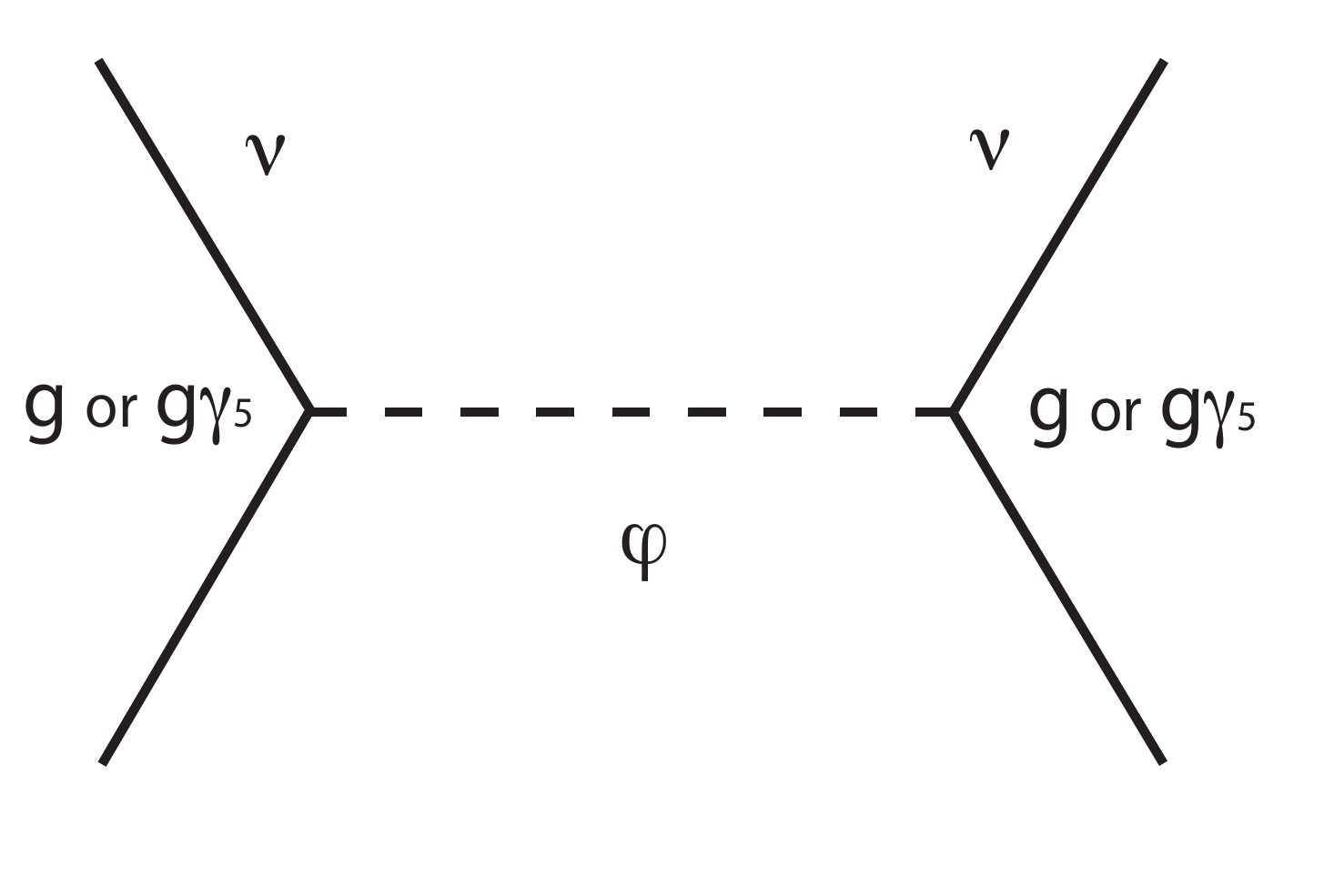}
\caption{$\nu$--$\nu$ long range forces induced by $\phi$ exchange}
\label{fig2}
\end{figure*}

The cosmological neutrino masses which we have explored depend crucially on the viability of the slow roll
regime for ultralight fields. To make the masses significant and delay their turn-off we need ultralight bosons with very large initial field displacements from their minima. The displacements should be either mildly super-Planckian or at least near-Planckian. In fact the proposals to identify quintessence with CP-violating phase of the neutrino mass matrix \cite{barbierihall,hill}, which assume the neutrino masses resulting from Yukawa couplings to such fields are constant, require $\phi \gg M_{\rm Pl}$. Axionic quintessence likewise requires
$\phi > M_{\rm Pl}$ to reproduce the dark energy equation of state sufficiently close to $-1$ and fit the data,
although some slow roll deviations can be allowed \cite{qbounds}. This might be problematic from the point of view of WGC \cite{nima}, that seems to limit super-planckian fields.

The conflict between having large field vevs to yield this type of neutrino masses, and mimic dark energy if
it is quintessence, which could all be tested experimentally, and the inferences from WGC
about aspects of quantum gravity raises an interesting possibility of neutrino cosmology as a test of quantum gravity. If we really observe time variation of neutrino masses, sterile neutrino fraction and dark energy, which are all correlated, we may end up identifying Planckian fields as the cause. Even finding fields which are sub-Planckian but very large would be interesting for this purpose, since it would point to the existence of potentials with very large periods, such as those encountered in monodromy models.

In summary, we pointed out that neutrinos can couple to ultralight bosons without spoiling the flatness of their potentials. If such ultralight fields are in slow roll they can give a mass to neutrinos. In our minimal setup the SM gauge symmetries require this to be the sterile neutrino mass. The scenario can naturally realize both the see-saw regime and the mixed neutrino regime, as well as interpolate between the two: see-saw at early times transitioning to mixed regime in a late universe. The evolution of the sterile neutrino mass might help to reconcile the lab experiments suggesting mixed sterile neutrinos with cosmological bounds disfavoring them, by avoiding equilibrated sterile neutrinos in the early universe. The scenario is not strongly constrained at present, as it is more circumspect than the approaches in \cite{barbierihall,hill,mavans}. Instead of trying to explain the origin of dark energy we are instead pointing out that ultralight dark scalars and neutrinos can consistently couple with scalar vevs yielding $\nu$ masses. In turn neutrinos could be a probe of various aspects of ultralight dark sectors, including dark energy. More work is needed to explore quantitative aspects of such models.

\section*{Acknowledgements}
We would like to thank Albion Lawrence, Martin Sloth and Alessandro Strumia for useful discussions. NK would like to thank CERN Theory Division and ${\rm CP}^3$ Origins, University of Southern Denmark, for kind hospitality in the course of this work. T.H. and N.K. are supported in part by the DOE Grant DE-SC0009999.


\vskip1cm


\begin{thebibliography}{99}


\bibitem{fhsw}
  J.~A.~Frieman, C.~T.~Hill, A.~Stebbins and I.~Waga,
  Phys.\ Rev.\ Lett.\  {\bf 75}, 2077 (1995).

\bibitem{nomura}
  Y.~Nomura, T.~Watari and T.~Yanagida,
  Phys.\ Lett.\ B {\bf 484}, 103 (2000).

   \bibitem{nilles}
  J.~E.~Kim and H.~P.~Nilles,
  Phys.\ Lett.\ B {\bf 553}, 1 (2003).

\bibitem{kalsor}
  N.~Kaloper and L.~Sorbo,
  JCAP {\bf 0604}, 007 (2006).

\bibitem{kalsor2}
  N.~Kaloper and L.~Sorbo,
  Phys.\ Rev.\ D {\bf 79}, 043528 (2009).

\bibitem{trivedi}
  S.~Panda, Y.~Sumitomo and S.~P.~Trivedi,
  Phys.\ Rev.\ D {\bf 83}, 083506 (2011).


\bibitem{quintax}
  G.~D'Amico, T.~Hamill and N.~Kaloper,
  Phys.\ Rev.\ D {\bf 94}, no. 10, 103526 (2016).


 \bibitem{barbierihall}
  R.~Barbieri, L.~J.~Hall, S.~J.~Oliver and A.~Strumia,
  Phys.\ Lett.\ B {\bf 625}, 189 (2005).

\bibitem{hill}
  C.~T.~Hill, I.~Mocioiu, E.~A.~Paschos and U.~Sarkar,
  Phys.\ Lett.\ B {\bf 651}, 188 (2007).

\bibitem{jaeckel}
  J.~Jaeckel, V.~M.~Mehta and L.~T.~Witkowski,
  JCAP {\bf 1701}, no. 01, 036 (2017).

\bibitem{DM}
  A.~Berlin,
  Phys.\ Rev.\ Lett.\  {\bf 117}, no. 23, 231801 (2016);
  A.~Berlin and D.~Hooper,
  Phys.\ Rev.\ D {\bf 95}, no. 7, 075017 (2017);
  Y.~Zhao,
  Phys.\ Rev.\ D {\bf 95}, no. 11, 115002 (2017);
  G.~Krnjaic, P.~A.~N.~Machado and L.~Necib,
  Phys.\ Rev.\ D {\bf 97}, no. 7, 075017 (2018);
  V.~Brdar, J.~Kopp, J.~Liu, P.~Prass and X.~P.~Wang,
  Phys.\ Rev.\ D {\bf 97}, no. 4, 043001 (2018);
  H.~Davoudiasl, G.~Mohlabeng and M.~Sullivan,
  arXiv:1803.00012 [hep-ph].



\bibitem{banksdine}
  T.~Banks, M.~Dine, P.~J.~Fox and E.~Gorbatov,
  JCAP {\bf 0306}, 001 (2003).

\bibitem{witten}
  P.~Svrcek and E.~Witten,
  JHEP {\bf 0606}, 051 (2006).

\bibitem{nima}
  N.~Arkani-Hamed, L.~Motl, A.~Nicolis and C.~Vafa,
  JHEP {\bf 0706}, 060 (2007).

\bibitem{qbounds}
  V.~Smer-Barreto and A.~R.~Liddle,
  JCAP {\bf 1701}, no. 01, 023 (2017);
  S.~Dhawan, A.~Goobar, E.~Mortsell, R.~Amanullah and U.~Feindt,
  JCAP {\bf 1707}, no. 07, 040 (2017).


\bibitem{langacker}
  P.~Langacker, ``The standard model and beyond,''
  Boca Raton, USA: CRC Pr. (2010).

\bibitem{judge}
  G.~F.~Giudice,
  arXiv:1710.07663 [physics.hist-ph].

\bibitem{amendolabarbieri}
  L.~Amendola and R.~Barbieri,
  Phys.\ Lett.\ B {\bf 642}, 192 (2006).

\bibitem{mavans}
  R.~Fardon, A.~E.~Nelson and N.~Weiner,
  JCAP {\bf 0410}, 005 (2004).

\bibitem{zaldarriaga}
  N.~Afshordi, M.~Zaldarriaga and K.~Kohri,
  Phys.\ Rev.\ D {\bf 72}, 065024 (2005).

  \bibitem{barbieridolgov}
  R.~Barbieri and A.~Dolgov,
  Phys.\ Lett.\ B {\bf 237}, 440 (1990);
  Nucl.\ Phys.\ B {\bf 349}, 743 (1991).

\bibitem{Langacker:1989sv}
  P.~Langacker,
  UPR-0401T.

\bibitem{murayama}
  A.~Pierce and H.~Murayama,
  Phys.\ Lett.\ B {\bf 581}, 218 (2004).

\bibitem{strumia}
  M.~Cirelli, G.~Marandella, A.~Strumia and F.~Vissani,
  Nucl.\ Phys.\ B {\bf 708}, 215 (2005).

\bibitem{hannes}
S. Hannestad, I. Tamborra, and T. Tram,
JCAP {\bf 7}, 025 (2012). 

\bibitem{white}
  K.~N.~Abazajian {\it et al.},
  arXiv:1204.5379 [hep-ph];
  K.~N.~Abazajian {\it et al.} [CMB-S4 Collaboration],
  arXiv:1610.02743 [astro-ph.CO];
M. Archidiacono, S. Gariazzo, C. Giunti, S. Hannestad, R. Hansen, M. Laveder, T. Tram,
JCAP {\bf 08}, 067 (2016). 
  %

\bibitem{moodywilczek}
 J.~E.~Moody and F.~Wilczek,
 Phys.\ Rev.\ D {\bf 30}, 130 (1984);
  F.~Ferrer and J.~A.~Grifols,
  Phys.\ Rev.\ D {\bf 58}, 096006 (1998);
  B.~A.~Dobrescu and I.~Mocioiu,
  JHEP {\bf 0611}, 005 (2006).

\bibitem{spin}
E. Fischbach and D. E. Krause,
Phys. Rev. Lett. {\bf 82}, 4753. 

\bibitem{stars}
  G.~G.~Raffelt,
  Ann.\ Rev.\ Nucl.\ Part.\ Sci.\  {\bf 49}, 163 (1999);
  J.~A.~Grifols, E.~Masso and S.~Peris,
  Mod.\ Phys.\ Lett.\ A {\bf 4}, 311 (1989).



\end{thebibliography}
\end{document}